
\magnification = 1200
\hsize = 15 truecm
\vsize = 23 truecm
\baselineskip 20 truept
\voffset = -0.5 truecm
\parindent = 1 cm
\overfullrule = 0pt
\count0 = 1
\footline={\hfil}
\def\D{D\kern -0.7em \hbox {$D$}}
\def\p{\partial\kern -0.7em \hbox {$\partial$}}

\null

\settabs 3 \columns
\+&&  Preprint-KUL-TF-93/39 \cr
\+&&  hep-th/9309103 \cr
\+&&  September 1993 \cr

\null
\vskip 1 truecm

\centerline
{\bf  Quantization of the topological $\sigma$-model and}
\centerline
{\bf  the master equation of the BV formalism}

\vskip 3 truecm

\centerline{\bf S. Aoyama}
\smallskip

\vskip 0.5 cm

\centerline{\bf Instituut voor Theoretische Fysica}
\smallskip
\centerline{\bf Katholieke Universiteit Leuven}
\smallskip
\centerline{\bf Celestijnenlaan 200D}
\smallskip
\centerline{\bf B-3001 Leuven, Belgium}

\vskip 2.5 truecm

\noindent
{\bf Abstract}

We quantize the topological $\sigma$-model. The quantum master equation of the
Batalin-Vilkovisky formalism $\Delta_\rho \Psi = 0$ appears as a condition
which eliminates  the exact states from the BRST invariant states  $\Psi$
defined by $Q \Psi = 0$. The phase space of the BV formalism is a supermanifold
with a specific symplectic structure, called the fermionic K\"ahler manifold.

\vskip 2 cm
\noindent
e-mail: shogo\%tf\%fys@cc3.kuleuven.ac.be

\vfill\eject

\footline={\hss\tenrm\folio\hss}

There is much interest in the BV formalism$^{[1\sim 9]}$.  Originally it was
devised to study quantization of gauge theories$^{[10,11]}$. The geometry of
the formalism was clarified in refs 1 and 4$\sim$6. Its extended viability has
been shown by the applications to the non-critical string$^{[3,7]}$ and the
string field theory$^{[2,5,8]}$.  But there we may recognize a conceptual
departure from the original framework of the BV formalism. The anti-bracket
does not follow directly from an action as defining the S-T identity$^{[12]}$.
Recently, as one of such approaches,  the BV  formalism  was  discussed on a
supermanifold with a specific symplectic structure, called the fermionic
K\"ahler manifold$^{[13]}$.
In this letter
we show that this fermionic K\"ahler geometry naturally appears by quantizing
the topological $\sigma$-model$^{[14]}$. Namely
the quantum master equation $\Delta_\rho \Psi = 0$ with the symplectic
structure of this geometry is obtained as a condition  which  eliminates  the
exact states of the BRST invariant states $\Psi$ defined by $Q \Psi = 0$.
In other words $\Delta_\rho$ operator is equivalent to the superpartner of the
BRST charge $Q$ in the topological $\sigma$-model.

The study in this direction has been motivated by the work by Lian and
Zuckerman$^{[7]}$. They  conjectured that in the topological conformal field
theory the $G_R $(or $G_L$) charge could be connected with the $\Delta_\rho$
operator of the BV formalism. Although any concrete connection  was not given,
 they wondered about the meaning of the quantum master equation in such a case.

The study has been also motivated by the arguments in ref. 15. They discussed
the BRST cohomolgy of the topological $\sigma$-model in the right-moving
sector, $Q_R \Psi = 0$ with $G_R \Psi = 0$ (or eqivalently in the left-moving
sector). However the $G_R$ charge does not become the $\Delta_\rho$ operator of
the BV formalism by itself, although this cohomology equals the BRST cohomology
defined by $(Q_L + Q_R) \Psi =0$ with $(G_L + G_R)\Psi = 0$. (Namely the
Dolbeault cohomology equals the de Rham cohomology on the K\"ahler manifold.)
So we carefully study the latter cohomology. It is shown  that the $G_L + G_R$
charge equals the $\Delta_\rho$ operator of the BV formalism on the fermionic
K\"ahler manifold. On an off-shell state the charge $Q_L + Q_L$ becomes an
exterior derivative $d$ of a differential form, while the charge $G_L + G_R$(or
the $\Delta_\rho$ operator) its adjoint $d^*$. The last statement is generally
expected in the topological field theory because of the $N=2$ supersymmetric
structure$^{[14,16,17]}$. Thus the quantum master equation of the BV formalism
is a necessary and sufficient condition to pick out a unique representative for
each BRST cohomology class. The physical states are given by harmonic
differential forms of the K\"ahler manifold.

Through our study of the BRST cohomology the meaning of the function $\rho$ in
the master equation $\Delta_\rho \Psi = 0$ is considerably clarified. It is a
quantum effect of operator ordering in defining the quantum $G$ charge and can
be renormalized in physical states. We also give a substantial argument to show
that the function $\rho$ itself is an element of the BRST cohomology class,
i.e, a physical state.

\vskip 1 cm

To start with, it is convenient to recall the basic formulae of the symplectic
geometry$^{[4\sim 6]}$. We consider
a $2D$ bosonic manifold $M$ parametrized by real coordinates $u^\alpha =
(u^1,u^2,\cdots,u^{2D}) $. Suppose that it has a symplectic structure given by
a non-degenerate 2-form
$$
\Omega = du^\beta \wedge du^\alpha \Omega_{\alpha \beta}   ,    \eqno (1)
$$
which is closed $ d\Omega = 0$.
In components these equations imply that $\Omega_{\alpha \beta}  =
-\Omega_{\beta \alpha}$ and
$$
  \partial_\alpha \Omega_{\beta \gamma} + \partial_\beta \Omega_{\alpha \gamma}
+ \partial_\gamma \Omega_{\alpha \beta} = 0 ,    \eqno (2)
$$
or equivalently $\Omega^{\alpha \beta} = -\Omega^{\beta \alpha}$ and
$$
\Omega^{\alpha \eta}\partial_\eta \Omega^{\beta \gamma} + \Omega^{\beta
\eta}\partial_\eta \Omega^{\gamma \alpha} + \Omega^{\gamma \eta}\partial_\eta
\Omega^{\alpha \beta} = 0.  \eqno (3)
$$
by inverting $\Omega_{\alpha \beta}$ as
$$
\Omega_{\alpha \beta} \Omega^{\beta \gamma} = \Omega^{\gamma
\beta}\Omega_{\beta \alpha} = \delta^\gamma_\alpha.  \eqno (4)
$$

We introduce real fermion coordinates $\psi^\alpha =
(\psi^1,\psi^2,\cdots,\psi^{2D})$ and think of
a $4D$ manifold ${\cal M}$ parametrized by  supercoordinates $s^i =
(u^1,u^2,\cdots,u^{2D}, \psi^1,\psi^2,$ $\cdots,\psi^{2D}) $. Assume that
the symplectic structure $\Omega_{\alpha \beta}$ is given by a generalized
2-form
$$
\omega = ds^j \wedge ds^i \omega_{ij}   ,               \eqno (5)
$$
and  $  d\omega = 0$.
In components these equations read
$$
(-)^{ik}  \partial_i \omega_{jk} + (-)^{ji} \partial_j \omega_{ki} +
(-)^{kj}\partial_k \omega_{ij} = 0 ,    \eqno (6)
$$
$$
\omega_{ij}  = -(-)^{ij}\omega_{ji}.  \eqno (7)
$$
Here we used the short-hand notation for the grassmannian parity of the
coordinates $\varepsilon (s^i) = i $ in the sign factor.
So long as the grassmannian parity is assigned as $\varepsilon (\omega_{ij}) =
i + j $,
the 2-form given by eq. (5) is bosonic and defines the ordinary symplectic
structure.
However if the opposite
grassmannian parity is assigned as
$
\varepsilon (\omega_{ij}) = i + j + 1,
$
then the 2-form (5) is fermionic. Eqs (6) and (7) become respectively
$$
(-)^{(i+1)(k+1)}\omega^{il}\partial_l \omega^{jk} +
(-)^{(j+1)(i+1)}\omega^{jl}\partial_l \omega^{ki} +
(-)^{(k+1)(j+1)}\omega^{kl}\partial_l \omega^{ij} = 0,
\eqno (8)
$$
$$
\omega^{ij} = -(-)^{(i+1)(j+1)}\omega^{ji}, \eqno (9)
$$
by inverting $\omega_{ij}$ as
$$
\omega_{ij} \omega^{jk} = \omega^{kj} \omega_{ji} = \delta^k_i.
$$
Then the anti-bracket is defined by
$$
\{A,B\} = A\overleftarrow \partial_i \omega^{ij} \partial_j B.  \eqno (10)
$$
The right-derivative $\overleftarrow \partial_i$ is related with the left-one
by
$$
A\overleftarrow \partial_i = (-)^{i(\varepsilon (A) + 1)} \partial_i A  .
$$
With this fermionic symplectic structure we define also a second order
differential operator by
$$
\Delta_\rho \equiv {1 \over \rho }(-)^i \partial_i(\rho \omega^{ij}\partial_j),
\eqno (11)
$$
where $\rho$ is a bosonic function of $y^i$. It is related with the
anti-bracket through
$$
\Delta_\rho (AB) = \Delta_\rho A \cdot B + (-)^{\varepsilon (A)} A\Delta_\rho B
 +
(-)^{\varepsilon (A) } \{A,B \}. \eqno (12)
$$
A crucial observation in this letter is that as a special solution to eqs (8)
and (9)  we have the fermionic symplectic structure
$$
\eqalignno{
\omega^{ij} & = \left(
\matrix{ \omega^{u u} & \omega^{u \psi}  \cr
         \omega^{\psi u} & \omega^{\psi \psi}  \cr } \right)  &  \cr
&       &     \cr
& = \left(
\matrix{ 0    & \Omega^{\alpha \beta}  \cr
 &      \cr
 \Omega^{\alpha \beta}   &    \psi^\gamma{\partial \over \partial u^\gamma}
\Omega^{\alpha \beta}    \cr}\right),  &  (13) \cr}
$$
in which $\Omega^{\alpha \beta}$ is the one of the bosonic submanifold $M$
obeying eq. (3).

\vskip 1 cm

If $M$ is a K\"ahler manifold, it is endowed with a complex structure which is
covariantly constant:
$$
D_\gamma J^\alpha_{\ \beta} = 0,  \eqno (14)
$$
and $J^\alpha_{\ \beta} J^\beta_{\ \gamma} = -\delta^\alpha_\gamma$.
We assume the metric $g_{\alpha \beta}$ to be of type $(1,1)$, i.e.,
$$
g_{\alpha \beta} = g_{\gamma \delta} J^\gamma_{\ \alpha} J^\delta_{\ \beta}.
$$
The symplectic structure $\Omega_{\alpha \beta}$ is given by
$$
\Omega_{\alpha \beta} = g_{\alpha \gamma} J^\gamma_{\ \beta}.
$$
The closure property (2) follows from eq. (14).
We also obtain the self-duality condition of the Riemann tensor
$$
R_{\alpha \beta \gamma \delta} = R_{\alpha \beta \eta \sigma}J^\eta_{\ \
\gamma}
J^\sigma_{\ \delta},  \eqno (15)
$$
taking the covariant derivative of eq. (14). According to the previous
discussion we may generalize the K\"ahler manifold $M$ to a supermanifold
${\cal M}$ with the fermionic symplectic structure given by eq. (13). In this
case the complex structure $J^\alpha_{\ \beta }$ may be also extended to
$J^i_{\ j}$ as
$$
J^i_{\ j} = \left(
\matrix{ J^\alpha_{\ \beta} &  0  \cr
         0  &  J^\alpha_{\ \beta} \cr}\right).
$$
It is natural to define the metric of the supermanifold ${\cal M}$ by
$$
\gamma_{ij} = - \omega_{ik} J^k_{\ j}.
$$
Then the metric is fermionic, $\varepsilon (\gamma_{ij}) = i+j+1 $ and
of type $(1,1)$.
It is given by
$$
\gamma_{a \underline b} = \partial_a \partial_{\underline b} K,$$
in which $K$ is a fermionic K\"ahler potential.
Here we have used the complexcified coordinates $u^\alpha = (u^a,u^{\underline
a}), \quad a,\ \underline a = 1,2,\cdots,D$. This supermanifold ${\cal M}$ is a
fermionic version of the K\"ahler manifold $M$, which has been discussed in
refs 13. It is called the fermionic K\"ahler manifold.

\vskip 1 cm

The topological $\sigma$-model$^{[14]}$ is constructed on the (bosonic)
K\"ahler manifold discussed just above. In the real field representation the
action is given by
$$
\eqalign{
S = \int d^2 x [-{1 \over 4} & H^{\mu \alpha}H_{\mu \alpha}\ + \
H^\mu_{\ \alpha} \partial_\mu u^\alpha  \cr
&  -i\rho^\mu_{\ \alpha} J^\alpha_{\ \beta} D_\mu \psi^\beta \ - \
{1 \over 8}R_{\alpha \beta}^{\ \ \ \gamma \delta} \rho^\mu_{\ \gamma} \rho_{\mu
\delta} \psi^\alpha \psi^\beta \  ]   \cr }   \eqno (16)
$$
Here we have spin $0$ fermions $\psi^\alpha$, spin $1$ fermions $\rho_{\mu
\alpha}$ and spin $1$ bosons $H^{\ \alpha}_\mu$ in addition to spin $0$ bosons
which are coordinates of the K\"ahler manifold $M$. The spin $1$ fields are
constrained by the self-duality conditions
$$
\rho^\mu_{\ \alpha} = \varepsilon^\mu_{\ \nu}J^\alpha_{\ \beta}\rho^{\nu
\beta},
\quad \quad \quad
H^\mu_{\ \alpha} = \varepsilon^\mu_{\ \nu}J^\alpha_{\ \beta}H^{\nu \beta}.
\eqno (17)
$$
Here $ \varepsilon^\mu_{\ \nu} $ is the anti-symmetric tensor in two
dimensions. By the replacement $\psi^\alpha = J^\alpha_{\ \beta} \chi^\beta$
and the use of eq. (15) the action (16) goes back to the original form in ref.
14.
It is invariant by the BRST transformations
$$
\eqalign{\delta u^\alpha & = \ \ i J^\alpha_{\ \beta}\psi^\beta,    \cr
\delta \psi^\alpha & = \ \ iJ^\alpha_{\ \beta, \gamma}(J^\beta_{\ \eta}
\psi^\eta)(J^\gamma_{\ \delta} \psi^\delta),   \cr
\delta \rho_{\mu \alpha} & = \ \ H_{\mu \alpha}\ - \ i \rho_{\mu
\beta}\Gamma^\beta_{\alpha \gamma}
(J^\gamma_{\ \delta}\psi^\delta),     \cr
\delta H^{\ \alpha}_\mu & = -i \Gamma^\alpha_{\beta \gamma}(J^\beta_{\
\eta}\psi^\eta)H^\gamma_\mu \ - \ {1 \over 2 } R_{\gamma \delta}^{\ \ \ \alpha
\beta}\rho_{\beta \mu} \psi^\gamma \psi^\delta,  }  \eqno (18)
$$
For the canonical formulation it is convenient to write the action (16) in
terms of the independent fields alone. In representation $\varepsilon^0_{\ 1} =
- \varepsilon^1_{\ 0} = 1$ and $\varepsilon^0_{\ 0} = \varepsilon^1_{\ 1} = 0$
the constraints (17) read
$$
\rho_{1 \alpha} = \rho_{0 \beta}J^\beta_{\ \alpha}, \quad\quad\quad
H_{1 \alpha} = H_{0 \beta}J^\beta_{\ \alpha}.
$$
By using these equations we eliminate the fields $\rho_{1 \alpha}$ and $H^{\
\alpha}_1$ from the action. It becomes
$$
\eqalign{
S = \int d^2 x [- & {1 \over 2} H^{\ \alpha}_0 H_{0 \alpha}\ + \
H_{0 \alpha} \partial_0 u^\alpha \ + \
H_{0 \alpha}J^\alpha_{\ \beta}\partial_1 u^\beta        \cr
&  -i\rho_{0 \alpha} J^\alpha_{\ \beta} D_0 \psi^\beta
 + i\rho_{0 \alpha}D_1 \psi^\alpha \ - \
{1 \over 4}R_{\alpha \beta}^{\ \ \ \gamma \delta} \rho_{0 \gamma} \rho_{0
\delta} \psi^\alpha \psi^\beta \  ],   \cr }
$$
by eq. (15).
We carry out the canonical formulation at the euclidean time $x_0 = 0$. The
standard Noether procedure gives the Hamiltonian and the BRST charge
$$
\eqalignno{
H & = \int d x_1 T_{00}  & \cr
& = \int d x_1 [\ {1 \over 2} H^{\ \alpha}_0 H_{0 \alpha}\ - \
H_{0 \alpha} J^\alpha_{\ \beta} \partial_1 u^\beta   & \cr
 & \quad \quad \quad \quad \quad \quad \quad \quad
- \ i\rho_{0 \alpha} D_1 \psi^\alpha \ + \
  {1 \over 4}R_{\alpha \beta}^{\ \ \ \gamma \delta} \rho_{0 \gamma} \rho_{0
\delta} \psi^\alpha \psi^\beta \  ] &  \cr
Q & = i\int d x_1 [\ (\ H_{0 \alpha} \ - \ i\rho_{0 \sigma}J^\sigma_{\ \eta}
\Gamma^\eta_{\alpha \beta}\psi^\beta)(J^\alpha_{\ \gamma}\psi^\gamma)   & \cr
&  \quad \quad \quad \quad \quad \quad \quad \quad \quad \quad \quad \quad
\quad
+ \ i\rho_{0 \sigma}J^\sigma_{\ \eta}J^\eta_{\ \alpha, \beta}
(J^\alpha_{\ \gamma}\psi^\gamma) (J^\beta_{\ \delta}\psi^\delta)].
  & (19) \cr}
$$
For these quantities we find that
$$
T_{00} = \delta G_{00},
$$
 with
$$
G_{00} = {1 \over 2 }g^{\alpha \beta}\rho_{0 \alpha}[H_{0 \beta} -
2\Omega_{\beta \gamma}
\partial_1 u^\gamma \ ],  \eqno (20)
$$
which is the hallmark of the topological field theory.

\vskip 1 cm

We calculate the canonical conjugate momenta for $u^\alpha$ and $\psi^\alpha$:
$$
\eqalign{
\pi^u_\alpha & = H_{0 \alpha} - i\rho_{0 \sigma}J^\sigma_{\
\gamma}\Gamma^\gamma_{\alpha \beta}\psi^\beta, \cr
\pi^\psi_\alpha & = i\rho_{0 \beta}J^\beta_{\ \alpha},  \cr}
$$
and set up the Poisson brackets as usual:
$$
\eqalign{
\{\pi^u_\alpha (0,x_1), u^\beta (0,y_1) \}_{PB} & = \delta^\beta_\alpha
\delta^{(1)} (x_1 - y_1),    \cr
\{\pi^\psi_\alpha (0,x_1), \psi^\beta (0,y_1) \}_{PB} & = \delta^\beta_\alpha
\delta^{(1)} (x_1 - y_1).    \cr}
$$
As a consistency check it is shown that the Hamiltonian $H$ and the $Q$ charge
are  generators of the Euler-Lagrange equations  and
the BRST transformations (18) respectively. Consequently it follows that
$$
\eqalignno{
\{Q, G_{00} \}_{PB} & = T_{00},  \cr
\{Q, Q \}_{PB} & = 0   &    \cr}
$$
Then we write the $Q$ charge and the space integral of $G_{00}$, called the $G$
charge, in terms of the canonical variables alone:
$$
\eqalignno{
Q & = i\int d x_1 [\pi^u_\alpha (J^\alpha_{\ \gamma}\psi^\gamma)
 \ + \ \pi_\eta J^\eta_{\ \alpha, \beta} (J^\alpha_{\ \gamma}\psi^\gamma)
 (J^\beta_{\ \delta} \psi^\delta)\ ] ,  & (21)\cr
G & = -{i \over 4}\int d x_1 [\Omega^{\alpha \beta}\pi^\psi_{[\alpha}
\pi^u_{\beta]}
\ - \ \Omega^{\alpha \beta}_{\ \ \ , \gamma}\psi^\gamma \pi^\psi_\alpha
\pi^\psi_\beta \ - \ 4 \pi^\psi_\alpha \partial_1 u^\alpha \ ].  & (22)  \cr}
$$
The last formula has been derived by using formula
$$
\Omega^{\alpha [\beta} \Gamma^{\gamma]}_{\alpha \delta} =
\Omega^{\beta \gamma}_{\ \ \ , \delta}.
$$

\vskip 1 cm

Now we are in a position to quantize the topological $\sigma$-model.
By the replacement
$$
\eqalign{
\pi^u_\alpha (0,y) & = {\delta \over \delta u^\alpha (0,y)} \equiv \p_\alpha
(0,y),
\cr
\pi^\psi_\alpha (0,y) & = {\delta \over \delta \psi^\alpha (0,y)} \equiv
\D_\alpha (0,y).
\cr }
$$
The $Q$ and $G$ charges become quantum operators. The quantum Hamiltonian
is obtained through $H = [Q,G]_+$. Physical states $\Psi$ of the theory are
defined by
$$
Q \Psi = 0.  \eqno (23)
$$
They are also assumed to obey the condition$^{[16]}$
$$
G \Psi = 0.   \eqno (24)
$$
It will be later shown that this extra requirement does a right thing.
Let us consider an off-shell  state given by
$$
\Psi (0,x) = \phi_{\alpha_1 \alpha_2 \cdots \alpha_N}(0,x) \psi^{\alpha_1}(0,x)
\psi^{\alpha_2} (0,x) \cdots \psi^{\alpha_N} (0,x), \quad \quad \quad N \le 2D.
\eqno (25)
$$
in which $\phi_{\alpha_1 \alpha_2 \cdots \alpha_N} (0,x)$ is a function of
$u^\alpha (0,x)$ with no derivative. Acting on this state the quantum $G$
charge drops
the last term of  the classical expression (22) in order to fulfil the
condition (24). Then the corresponding classical $G$ charge satisfies
$\{G,G\}_{PB} = 0$. Therefore we require that the quantum $G$ charge is
nilpotent as well as the $Q$ charge.
Eqs  (23) and (24) respectively obtain the following normal-ordered
expressions
$$
\eqalignno{
Q \Psi (0,x) & = i\int d y [J^\alpha_{\ \beta} \psi^\beta \p_\alpha \ + \
J^\eta_{\ \alpha, \beta} (J^\alpha_{\ \gamma} \psi^\gamma)(J^\beta_{\
\delta}\psi^\delta)\D_\eta \ ]\Psi (0,x) = 0,     &  (26) \cr
G \Psi (0,x) & = -{i \over 4}\int d y [\Omega^{\alpha \beta}\D_{[\alpha}
\p_{\beta]}
\ - \ \Omega^{\alpha \beta}_{\ \ \ , \gamma}\psi^\gamma \D_\alpha \D_\beta ]
\Psi (0,x)
 = 0.  & (27)   \cr}
$$
It is worth checking that in this operator realization the $Q$ and $G$ charges
indeed satisfy
$$
Q^2 = 0 \quad \quad {\rm and} \quad \quad  G^2 = 0.  \eqno (28)
$$
Remarkably eq. (27) can be put in a geometrical form of the BV formalism  such
that
$$
G_\rho \Psi (0,x) = -{i \over 4}\Delta_\rho \Psi (0,x) = 0,  \eqno (29)
$$
with  $\Delta_\rho$ defined by eq. (11), in which $\omega^{ij}$ is the
symplectic strucutre  of the fermionic K\"ahler manifold  given
by eq. (13) and $\partial_i = (\p_\alpha (0,y), \D_\alpha (0,y))$. Here we
should understand the index $i$ in the formula (11) as standing also for the
space coordinate $y$.
A little calculation shows that $$
G_\rho \Psi (0,x) = G \Psi (0,x) \ - \ {i \over 4}\delta^{(1)}(0)\{\log \rho
(0,x), \Psi (0,x) \}.  \eqno (30)
$$
Here the second piece is the anti-bracket given by (10), which represents
a quantum effect due to operator ordering.
In front of it we have the well-known quantum singularity of the $\Delta_\rho$
operator$^{[9]}$ (but the one-dimensional $\delta$-singularity in our canonical
formulation).
Therefore the representation (29) is more general than (27).
In order to ensure that $G_\rho$ also satisfies eq. (28) it is sufficient to
impose the condition on $\rho$\ $^{[4]}$
$$
G \rho (0,x) = 0,   \eqno (31)
$$
for the r.h.s. of eq. (30) is calculated  as
$$
G_\rho \Psi (0,x) = {1 \over \rho (0,x)}G[\rho (0,x) \Psi (0,x) ] - {1 \over
\rho (0,x)}[G\rho (0,x) ]\cdot \Psi (0,x),   \eqno (32)
$$
by eq. (12).
Thus we have shown that the condition (29) becomes the quantum master equation
of the BV formalism together with the nilpotency condition (31). The relevant
phase space of the BV formalism is the fermionic K\"ahler manifold discussed
in refs 13.

\vskip 1 cm

Our final task is to solve eqs. (26) and (29). As can be seen from eq. (32) the
quantum effect of operator ordering is a multiplicative renormalization of the
state. The function $\rho$ should be further constrained by
$$
Q\rho (0,x) = 0,  \eqno (33)
$$
in order that both equations become
$$
Q \Psi_\rho (0,x) = 0, \quad \quad \quad G \Psi_\rho (0,x) = 0,  \eqno (34)
$$
with the renormalized state $\Psi_\rho = \rho \Psi$. Hence $\rho$ can be
regarded as a physical state as well. By eq. (25) we may put the renormalized
state in the form
$$
\Psi_\rho (0,x) = A_{\alpha_1 \alpha_2 \cdots \alpha_N} (J^{\alpha_1}_{\
\beta_1} \psi^{\beta_1})(J^{\alpha_2}_{\ \beta_2}
\psi^{\beta_2})\cdots(J^{\alpha_N}_{\ \beta_N} \psi^{\beta_N}) ,
$$
with a new function $A_{\alpha_1 \alpha_2 \cdots \alpha_N}$. Here the
$x$-dependence was  not written explicitly. Then the first equation in  (34)
becomes
$$
\partial_{\alpha_0} A_{\alpha_1 \alpha_2\cdots \alpha_N} \ \pm \ {\rm cyclic \
permutations} = 0,
$$
while the second one
$$
D^{\alpha_1} A_{\alpha_1 \alpha_2\cdots \alpha_N}  = 0,   \eqno (35)
$$
with the covariant derivative for the K\"ahler manifold. Thus the respective
operators $Q$ and $G$ turn into  an exterior derivative $d$ and its adjoint
$d^*$ of the differential $N$-form $A$. Solutions to these equations are given
by harmonic differential forms of the K\"ahler manifold. This result proves
that it was right to impose the condition (24).

\vskip 1 cm

In this letter we have quantized the topological $\sigma$-model by the
canonical formalism. The BRST cohomology of the physical states defined by $Q
\Psi = 0$ has been worked out. We have imposed the condition $G \Psi = 0$ and
shown that it is equivalent to the quantum master equation of the BV formalism.
The relevant phase space is the fermionic K\"ahler manifold discussed in refs
13. As generally expected in the topological field theory$^{[14,16,17]}$, the
conditions $Q \Psi = 0$ and $G \Psi =0$
are written by an exterior derivative $d$ of a differential form and its
adjoint $d^*$. Thus the quantum master equation is a necessary and sufficient
condition to eliminate exact states of the BRST cohomology. The physical states
of the topological $\sigma$-model are given by harmonic differential forms of
the K\"ahler manifold.
  Our discussions have considerably clarified the meaning of the function
$\rho$ in the quantum master equation. Namely it takes into account a quantum
effect of operator ordering in defining the quantum  $G$ charge. The physical
states $\Psi$ get a multiplicative normalization from this quantum effect such
that $\Psi_\rho = \rho \Psi$. We have also found that the function $\rho$
itself is an element of the BRST cohomology by eqs (31) and (33).

  Finally the reader might worry about the $\delta$-singularity of the $G$
operator which appeared in the calculations like eq. (30) and (35). The origin
of this divergence can be traced back to the Hamiltonian through the relation
$H = [Q,G]_+$.  It contains the same singularity. This is the usual fenomenon
in the Schr\"odinger representation of the quantum field theory.
Also he might worry about the quantum singularity in defining a local state
such as eq. (25). In this regard the situation is worse in the canonical
quantization in the right(or left)-moving frame$^{[15]}$. In that case the
coordinates of the K\"ahler manifold no longer commute with each other.
Consequently a product of the coordinates at the same space-time point is more
divergent than in our case. However because of
the global topological nature of the theory there might be a circumstance in
which such singularities do not come in.
In mathematical words there could exist a Hilbert subspace in which
both $Q$ and $G$ charges  are realized merely  by the usual derivatives instead
of the functional ones.
 Note that the topological $\sigma$-model has instanton solutions. It is
promising to find  such a situation by  a semi-classical quantization around an
instanton background. It is interesting to demonstrate this conjecture.

\vskip 1 cm

\noindent
Acknowledgments

 The author is grateful to A. Van Proeyen for the discussions and reading the
manuscript. He thanks the Research Council of K.U. Leuven for the financial
support.

\vskip 3 cm

\noindent
{\bf References}

\noindent
\item{1.} E. Witten, Mod. Phys. Lett. A5(1990)487.
\item{2.} B. Zwiebach, Nucl. Phys. B390(1993)33.
\item{3.} E. Verlinde, Nucl. Phys. B381(1992)141.
\item{4.} A. Schwarz, Comm. Math. Phys. 155(1993)249.
\item{5.} E. Witten, ``On background independent open-string field theory",
IASSNS-HEP-92/53, hep-th/9208027, August 1992.
\item{6.} I. A. Batalin and I. V. Tyutin, Intern. J. Mod. Phys. A8(1993)2333.
\item{7.} B. Lian and G. Zuckerman, Comm. Math. Phys. 153(1993)613.
\item{8.} H. Hata and B. Zwiebach, ``Developing the covariant
Batalin-Vilkovisky approach to string theory", MIT-CTP-2177, hep-th/9301097,
January 1993.
\item{9.} W. Troost, P. van Nieuwenhuizen and A. Van Proeyen, Nucl. Phys.
B\-333  (1990)\-727;
\item{} A. Van Proeyen and S. Vandoren, ``Simplifications in Lagrangian BV
quantization exemplified by the anomalies of chiral $W_3$ gravity",
hep-th/9306147.
\item{10.} I.A. Batalin and G. A. Vilkovisky, Phys. Lett. B102(1981)27; Phys.
Rev. D28(1983)2567.
\item {11.}M. Henneaux and C. Teitelboim, ``Quantization of gauge systems",
Princeton University Press, Princeton, New Jersey, 1992.
\item{12.} J. Zinn-Zustin, Springer Lecture Notes in Physics 37(1975)1.
\item{13.} S. Aoyama and S. Vandoren, ``The Batalin-Vilkovisky formalism on
ferm\-ionic K\"ahler manifolds", KUL-TF-93/15, hep-th/9305087, May 1993;
\item{} S. Aoyama, ``The Batalin-Vilkovisky formalism with the Virasoro
symmetry", KUL-TF-93/16, May 1993;
\item{} O.M. Khudaverdian and A.P. Nersessian, ``On the geometry of the BV
formalism", UGVA-93/03-807, March 1993.
\item{14.} E. Witten, Comm. Math. Phys. 118(1988)411.
\item{15.} R. Dijkgraaf, H. Verlinde and E. Verlinde, ``Notes on topological
string theory and $2D$ quantum gravity", PUPT-1217, IASSNS-HEP-90/80,
November 1990.
\item{16.} W. Lerche, C. Vafa and N.P. Warner, Nucl. Phys. B324(1989)427.
\item{17.} E. Witten, Comm. Math. Phys. 117(1988)353;
\item{} T. Eguchi and S.K. Yang, Mod. Phys. Lett. A5(1990)1693.

\bye